\documentclass[prx,twocolumn,english,superscriptaddress,longbibliography,floatfix]{revtex4-2}
\usepackage{amsmath,amsfonts,amssymb,bm}
\usepackage{wrapfig}
\usepackage{graphicx}
\usepackage{subfigure}
\usepackage{appendix}
\usepackage{bbm}
\usepackage{dsfont}
\usepackage{ulem}
\usepackage{color}
\usepackage{physics} 
\usepackage{natbib}
\usepackage{etoolbox}
\usepackage{lineno}
\usepackage{mathrsfs}
\usepackage[colorlinks,linkcolor=blue,anchorcolor=green,citecolor=blue]{hyperref}
\usepackage{bookmark}

\begin{document}
\setlength{\columnsep}{0.8cm}
\title{Efficient Detection of Preparing Quantum Remote States\\Using Coherence Quantum Benefits}
\date{\today}
\author{Yuan-Sung Liu}
\affiliation{Department of Engineering Science, National Cheng Kung University, Tainan 70101, Taiwan}
\affiliation{Center for Quantum Frontiers of Research $\&$ Technology, NCKU, Tainan 70101, Taiwan}

\author{Shih-Hsuan Chen}
\affiliation{Department of Engineering Science, National Cheng Kung University, Tainan 70101, Taiwan}
\affiliation{Center for Quantum Frontiers of Research $\&$ Technology, NCKU, Tainan 70101, Taiwan}

\author{Bing-Yuan Lee}
\affiliation{Department of Engineering Science, National Cheng Kung University, Tainan 70101, Taiwan}
\affiliation{Center for Quantum Frontiers of Research $\&$ Technology, NCKU, Tainan 70101, Taiwan}

\author{Chan Hsu}
\affiliation{Department of Engineering Science, National Cheng Kung University, Tainan 70101, Taiwan}
\affiliation{Center for Quantum Frontiers of Research $\&$ Technology, NCKU, Tainan 70101, Taiwan}

\author{Guang-Yin Chen}
\affiliation{Department of Physics, National Chung Hsing University, Taichung 40227, Taiwan}

\author{Yueh-Nan Chen}
\email{yuehnan@mail.ncku.edu.tw}
\affiliation{Center for Quantum Frontiers of Research $\&$ Technology, NCKU, Tainan 70101, Taiwan}
\affiliation{Department of Physics, National Cheng Kung University, Tainan 70101, Taiwan}

\author{Che-Ming Li}
\email{cmli@mail.ncku.edu.tw}
\affiliation{Department of Engineering Science, National Cheng Kung University, Tainan 70101, Taiwan}
\affiliation{Center for Quantum Frontiers of Research $\&$ Technology, NCKU, Tainan 70101, Taiwan}
\affiliation{Center for Quantum Science and Technology, Hsinchu 30013, Taiwan}

\begin{abstract}
A sender can prepare a quantum state for a remote receiver using preshared entangled pairs, only the sender's single-qubit measurement, and the receiver's simple correction informed by the sender. It provides resource-efficient advantages over quantum teleportation for quantum information. Here, we propose the most efficient approach to detect the remote state preparation (RSP) based on the quantum benefits powered by quantum coherence's static resources of the shared pairs and the dynamic resources both the RSP participants input. It requires only the receiver's \textit{minimum} of one additional coherence creation operation to verify RSP. Experimentally, we implement the introduced RSP assessment using different photon pair states generated from a high-quality polarization Sagnac interferometer, confirming the necessary and sufficient role played by the static and dynamic quantum coherence resources and demonstrating efficient RSP verification. Our results provide a route to efficiently assess RSP in practical scenarios such as quantum information in quantum networks.
\end{abstract}

\maketitle

\section{Introduction}\label{intro}
Quantum information processing is based on the fact that quantum benefits can help process classically impossible or less efficient information. Remote state preparation (RSP)~\cite{pati2000minimum,bennett2001remote} is one of the essential quantum-information tasks applied to applications prepared for quantum networking, such as the deterministic generation of single-photon states~\cite{jeffrey2004towards}, the preparation of single-photon hybrid entanglement~\cite{barreiro2010remote}, the initialization of atomic quantum memory~\cite{rosenfeld2007remote}, and facilitation of teleportation between atomic-ensemble nodes~\cite{bao2012quantum}. Moreover, RSP serves as a faithful quantum channel essential to assist in performing client-server blind quantum computation~\cite{gustiani2021blind}. The foundational principles inherent in RSP also contribute to measurement-based quantum information processes, including one-way quantum computing \cite{raussendorf2001one,you2007efficient,tanamoto2009efficient,wang2010robust} and entanglement-enabled networks \cite{pirker2018modular,wehner2018quantum}. 

In contrast to quantum teleportation~\cite{bennett1993teleporting}, another type of quantum communication task that needs Bell-state (entangling) measurements for transmitting an unknown qubit and costs two-bit transfer, the RSP protocol only requires the local measurements on the sender's qubit of an entangled pair shared with the receiver and one bit sent to inform the receiver for simple one-qubit correction. This resource-efficient feature makes RSP exceptionally attractive and more suitable for experiments under various quantum-information scenarios, as reviewed above~\cite{jeffrey2004towards,barreiro2010remote,rosenfeld2007remote,bao2012quantum,gustiani2021blind,raussendorf2001one,you2007efficient,tanamoto2009efficient,wang2010robust,pirker2018modular,wehner2018quantum}.

Owing to the advantages of RSP, evaluating the performance of this task and finding the primary resources behind RSP have become crucial problems. Recently, various methods have been proposed to examine the static resources of the shared pair~\cite{dakic2012quantum,horodecki2014can,kanjilal2018remote,nikaeen2020optimal,li2022correlated}. First, it finds that quantum discord~\cite{hu2018quantum} is a resource~\cite{dakic2012quantum} instead of entanglement. The geometric discord~\cite{dakic2010necessary} is the same as the RSP payoff function (or payoff function for short), which is used to evaluate the performance of RSP. However, some research has different views on which resource plays the most significant role in RSP. As shown in Ref.~\cite{horodecki2014can}, after optimizing the receiver's correction, there is no chance that a separable state can beat entanglement as a resource in RSP. More recently, Ref.~\cite{kanjilal2018remote} considers average performance a concern instead of using the worst-case scenario. This approach shows that the correlation matrix~\cite{correma} of the shared state is vital in RSP, whose payoff function is zero only for the product state. 

To seek which ultimate static resource of the shared pair is needed in RSP, it has been pointed out that in a fully optimized scenario, which includes the payoff maximization over the sender's measurement and meaningful maximization over the receiver's measurement and correction, the performance of RSP scales with the sum of the two largest eigenvalues of the squared correlation matrix~\cite{nikaeen2020optimal}. Not long before, it was shown that correlated coherence~\cite{corre} is a resource of RSP~\cite{li2022correlated}, in which their payoff function is related to the $l_{2}$-norm measure of correlated coherence~\cite{streltsov2017colloquium}. At the same time, their definition of an incoherent state is the maximally mixed state.

\begin{figure*}[t] 
\centering
\includegraphics[width=1\textwidth]{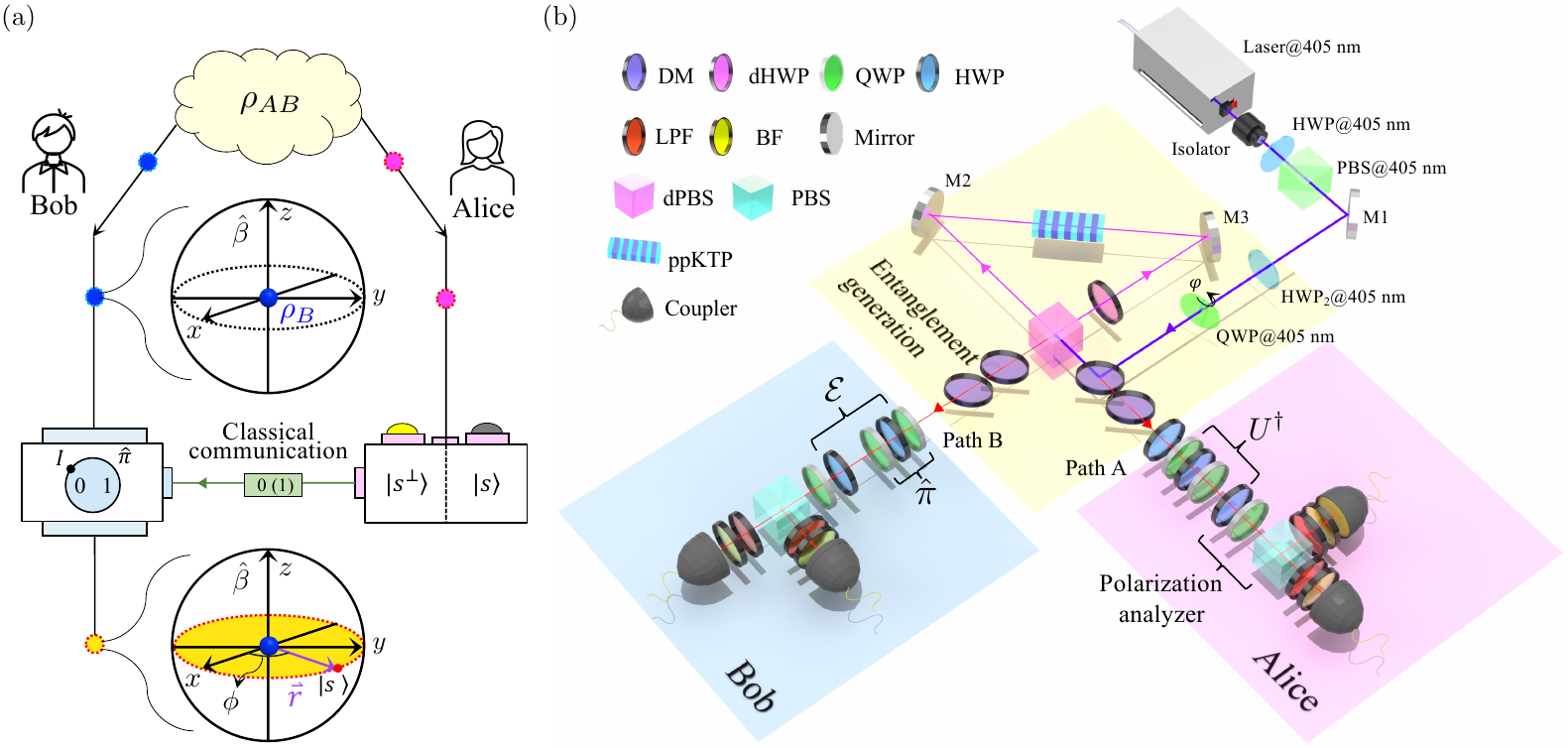}
\caption{Experimental setup for verifying RSP. (a) The RSP protocol. With the shared pairs, $\rho_{AB}=\ketbra{\psi^-}$, Alice aims to help Bob prepare an equatorial state $\ket{s}$ on Bob's Bloch sphere. First, Alice performs the prescribed measurement and sends the result to Bob in one way. Then, Bob receives a quantum state of his qubit, represented as a Bloch vector $\vec{r}$, after a $\hat{\pi}$ rotation, which depends on his received classical bit, 0 or 1. See Appendix~\ref{a} for the RSP implementation details. As illustrated and highlighted, the state on the first Bloch sphere represents Bob's state, $\rho_{B}$, without following the RSP protocol. The state on the second Bloch sphere represents the state received by Bob according to the standard RSP procedure. The yellow region increment emphasizes the quantum beneficial effect via RSP, enhancing the coherence of Bob's qubit state, $\rho_{B|A}=\ketbra{s}$. (b) Experimental RSP verification. The photon pairs generated from a PSI entangled photon source are close to the target state $\ket{\psi^{-}}=(\ket{H}_A\ket{V}_B-\ket{V}_A\ket{H}_B)/\sqrt{2}$. The logical qubit states for the $H$ and $V$ polarization states in the path $k$ are defined as: $\ket{H}_k:=\ket{0}$ and $\ket{V}_k:=\ket{1}$ for $k=A,B$, respectively. See Appendix~\ref{b} for the source details. The photon pairs are analyzed using the polarization analyzer consisting of half-wave plate (HWP), quarter-wave plate (QWP), and polarizing beam splitter (PBS) and are then collected by couplers with narrowband filters (BFs) and long-pass filters (LPFs) for subsequent detection by single photon counting modules (not shown). Alice’s $U^\dag$ and Bob’s verifying operations $\mathcal{E}$ are implemented by using a setting composed of a QWP, a HWP, and a QWP (QHQ) and a HWP in the polarization analyzer. Bob's correction operation $\hat{\pi}$ is alos implemented by the QHQ. Their experimental details can be found in Appendix~\ref{c}.}
\label{exptset}
\end{figure*}

In the practical situation, after transmitting photon pairs via quantum channel, the state may become any shared state that is arbitrarily impaired. Therefore, if we consider the scenarios shown in Refs.~\cite{dakic2012quantum} and~\cite{kanjilal2018remote}, according to their analysis, the states in their discussion are only for the sender's measurement direction parallel to the eigenvector with the maximum eigenvalue of the correlation matrix, which isotropically state are satisfied. Reference~\cite{kanjilal2018remote} uses an average method to judge the RSP performance. However, the resources state they can evaluate are almost the same as those in Ref.~\cite{dakic2012quantum}. Reference~\cite{nikaeen2020optimal} introduces the optimal exploitation source of RSP, which can apply to any state. However, as discussed above, their method requires the sender and receiver to be able to use any quantum operation.

As reviewed above, according to their evaluating and analysis methods, the sender and receiver must employ an optimized RSP protocol, where quantum state tomography (QST)~\cite{nielsen2010quantum} is needed to decompose their shared pairs to know the state density operator. It far goes against the simple requirements in the original RSP protocol. When implementing QST in the RSP scenario, it becomes clear that the receiver should be capable of performing three complementary single-qubit measurements. Also, some research~\cite{nikaeen2020optimal,li2022correlated} requires the receiver to apply any unitary operation for full optimization. Such high ability that a receiver should possess far exceeds the assumption underlying the RSP's original purpose: the sender can remotely prepare a state the ability-limited receiver can not. In addition to the complementary measurements, implementing QST needs two-way classical communication. This differs from the simple one-way classical communication in the standard RSP scenario, where the receiver is unnecessary to send classical information back to the receiver. These required abilities and operations, in addition to the needs in the standard RSP, can limit the usage and circumstances RSP can provide. Ultimately, they may restrict RSP's advantages and related quantum-information applications RSP can apply~\cite{jeffrey2004towards,barreiro2010remote,rosenfeld2007remote,bao2012quantum,gustiani2021blind,raussendorf2001one,you2007efficient,tanamoto2009efficient,wang2010robust,pirker2018modular,wehner2018quantum}.

Motivated by the above issue, we study what additional minimum resources are needed to evaluate RSP under the assumptions made in the standard RSP protocol. To tackle this problem, we develop a new concept and method to assess RSP based on the quantum benefits revealed by the static resources of the shared pairs and the dynamic resources both the RSP participants input. We show that the beneficial quantum impact RSP can provide originates from quantum coherence's static and dynamic resources. From the task-oriented viewpoint to consider the quantum benefits, these resources play a necessary and sufficient role in the RSP payoff for preparing quantum remote states, called the quantum-benefit RSP payoff. Furthermore, from the point of view of enhancing the quantum coherence via RSP to consider the quantum benefits, verifying quantum RSP by detecting the quantum benefits of coherence enhancement (or called coherence quantum benefits for short) only requires the receiver's one additional coherence operation. This is the minimum requirement needed other than the standard RPS. Moreover, we experimentally demonstrate the introduced RSP verification by using different photon states generated from a high-quality polarization Sagnac interferometer (PSI)~\cite{kim2006phase,fedrizzi2007wavelength,motazedifard2021nonlocal}. These implementations faithfully illustrate the RSP verification via the quantum-benefit RSP payoff for preparing quantum remote states and the efficient RSP verification based on the coherence quantum benefits.

The following introduces the quantum-benefit RSP payoff for preparing quantum remote states in Sec.~\ref{payoff}. In Sec.~\ref{iff}, we show that quantum coherence's static and dynamic resources are necessary and sufficient for the quantum-benefit RSP payoff. With these results as a basis, verifying RSP can be efficiently done by detecting the coherence quantum benefits with a minimum experimental requirement, as introduced in Sec.~\ref{coherenceqb}. The experimental verification of RSP based on the introduced quantum-benefit-based RSP evaluation concept and method is illustrated in Sec.~\ref{expt} afterward. The conclusion and outlook are given in Sec.~\ref{co}. Moreover, Appendices~\ref{a}-\ref{c} describe the technical details of the theoretical and experimental results.

\section{Quantum-benefit RSP payoff}\label{payoff}

We first consider a standard RSP scenario where a sender, Alice, wants to help a receiver, Bob, prepare a qubit in a maximal superposition state. According to the RSP protocol~\cite{pati2000minimum,bennett2001remote}, they initially share rotationally symmetric entangled Einstein-Podolsky-Rosen (EPR) pairs of the form:
\begin{equation}
\ket{\psi^{-}}=\frac{1}{\sqrt{2}}(U\ket{s_{0}} \otimes U\ket{s_{0}^{\perp}}-U\ket{s_{0}^{\perp}} \otimes U\ket{s_{0}}),\nonumber
\end{equation}
where $\ket{s_{0}}$ and $\ket{s_{0}^{\perp}}$ are arbitrary orthonormal states, and $U$ is a single-qubit unitary operator. After performing a measurement in the basis $\{U\ket{s_{0}},U\ket{s_{0}^{\perp}}\}$ by Alice, Bob's qubit becomes: $\ket{s}=U\ket{s_0}$ or $\ket{s^{\perp}}=U\ket{s^{\perp}_0}$. Conditioned on Alice's measurement result, she sends a classical bit telling Bob whether he needs to apply a $\pi$ rotation on his qubit along a direction denoted as $\hat{\beta}$. As illustrated in Fig.~\ref{exptset}(a), Alice aims to prepare a state with maximum coherence for Bob: $U\ket{s_{0}}=(\ket{0}+e^{i\phi}\ket{1})/\sqrt{2}$, where $\{\ket{0},\ket{1}\}$ is an orthonormal basis. This state can be characterized by a Bloch vector: $\vec{r}$, with $|\vec{r}|=1$, on the equator of the Bloch sphere. Here, Bob's $\pi$ rotation around the $z$-axis ($\hat{\beta}$), $\hat{\pi}$, is equal to the Pauli-$Z$ operation: $Z=\ketbra{0}-\ketbra{1}$. It is worth noting that Alice's measurement capability includes her ability to implement an unitary operation, $U^\dag$, related to the coherence creation operation $U$ such that $U\ket{s_{0}}$ becomes the target superposition state. See the detailed descriptions in Appendix~\ref{a}.

Compared to the ideal case in the RSP protocol described above, experimental RSP tasks are always imperfect and noisy. The existing evaluation methods mainly focus on the fidelity between the resulting Bob's state and the target state designed by Alice and define quadratic fidelity as the payoff function to evaluate the performance of RSP. However, as indicated in Sec.~\ref{intro}, their optimized protocol's required experimental can not fit the standard RSP scenario. Considering the effect of all participants' roles and shared pairs played in RSP [Fig.~\ref{exptset}(a)], we propose a different way of appraising the performance of RSP. We make the point on the difference between fidelity with and without following the RSP protocol. Such payoff function can directly reflect all the required resources in the RSP process: quantum coherence's static and dynamic resources required in the RSP.
 
Let us consider a single run of RSP process according to the standard RSP protocol. To continue Alice's assignment considered in the RSP task summarized above, she aims to prepare the state: $\ket{s}= (\ket{0}+ e^{i\phi}\ket{1})/\sqrt{2}$, orthogonal to the $z$-axis ($\hat{\beta}$) in the Bloch sphere [Fig.~\ref{exptset}(a)]. Here, Alice and Bob share qubit pairs with a state $\rho_{AB}=\sum_{m,n}\rho_{mn}\ketbra{m}{n}$ of the following general matrix form:
\begin{equation}
\rho_{AB}=\left[
\begin{matrix}
\rho_{11} & \rho_{12} & \rho_{13} & \rho_{14}\\
\rho_{21} & \rho_{22} & \rho_{23} & \rho_{24}\\
\rho_{31} & \rho_{32} & \rho_{33} & \rho_{34}\\
\rho_{41} & \rho_{42} & \rho_{43} & \rho_{44}\\
\end{matrix}\right],\label{rho}
\end{equation}
where $\ket{m}=\ket{m_A}\ket{m_B}$ for $m_A,m_B\in\{0,1\}$. Following the standard RSP protocol (see Appendix~\ref{a}), after Alice's perfect measurement (where $U$ is ideal as well) and Bob's ideal correction with respect to the state $\rho_{AB}$, we get the resulting state of Bob's qubit, denoted as $\rho_{B|A}$, where
\begin{equation}
\rho_{B\mid A}=\left[
\begin{matrix}
\rho_{11}+\rho_{33} & -e^{i\phi}\rho_{41}-e^{-i\phi}\rho_{23}\\
-e^{i\phi}\rho_{32}-e^{-i\phi}\rho_{14} & \rho_{22}+\rho_{44}\\
\end{matrix}\right]. \label{rhob}
\end{equation}
Meanwhile, the state derived without following the RSP protocol can be obtained by: $\rho_{B}={\rm tr}(\rho_{AB})$, and we have
\begin{equation}
\rho_{B}=\left[
\begin{matrix}
\rho_{11}+\rho_{33} & \rho_{12}+\rho_{34}\\
\rho_{21}+\rho_{43} & \rho_{22}+\rho_{44}\\
\end{matrix}\right].\label{rb}
\end{equation}
These resulting states of Bob's qubits can be evaluated by examining their similarities to the target state $\ket{s}$ in terms of state fidelity by:
\begin{eqnarray}
F(\rho_{B|A})&=& \text{tr}(\rho_{B\mid A}\rho_{s}) \nonumber\\
&=& \frac{1}{2}[1\!-\!(\rho_{32}+\rho_{23}+e^{2i\phi}\rho_{14}+e^{-2i\phi}\rho_{41})],
\end{eqnarray}
and
\begin{eqnarray}
F(\rho_{B})&=& \text{tr}(\rho_{B}\rho_{s}) \nonumber\\
&=& \frac{1}{2}[1+e^{i\phi}(\rho_{21}+\rho_{43})+e^{-i\phi}(\rho_{12}+\rho_{34})].
\end{eqnarray}

We define our \textit{quantum-benefit RSP payoff} function in a single run RSP as:
\begin{eqnarray}
\mathcal{P}&\equiv& F(\rho_{B|A})-F(\rho_{B})\nonumber \\ 
&=&-\frac{1}{2}[\rho_{32}+\rho_{23}+e^{2i\phi}\rho_{14}+e^{-2i\phi}\rho_{41}\label{qbpayoff}\\ \nonumber
&&\ \ \ \ \ -e^{i\phi}(\rho_{21}+\rho_{43})-e^{-i\phi}(\rho_{12}+\rho_{34})].
\end{eqnarray}
Equation~(\ref{qbpayoff}) can be used to reveal the importance of implementing the RSP protocol. That is, if the value of the quantum-benefit RSP payoff function is positive, $\mathcal{P}>0$, it means that Alice does help Bob prepare the target state using the shared pair state $\rho_{AB}$, showing the quantum benefit of RSP. On the other hand, if the payoff function is zero or negative, then Alice cannot afford any help via $\rho_{AB}$.

Moreover, since the target state lies on the equator of the Bloch sphere, the mean quantum-benefit RSP payoff averaged over $\mathcal{P}$ from $\phi=0$ to $\phi=2\pi$ (corresponding to all the states on the equator) is given by
\begin{eqnarray}
\mathcal{P}_{\text{avg}}&=&\frac{1}{2\pi}\int_{0}^{2\pi} \mathcal{P}\ d\phi\nonumber\\
&=&-\frac{1}{2} (\rho_{23}+\rho_{32}).
\label{pay}
\end{eqnarray}
This indicates a coherence criterion for the shared pair state's global coherence terms of $\rho_{AB}$, which enables Alice and Bob to show the RSP quantum benefit, $\mathcal{P}_{\text{avg}}>0$, on average.

\section{Necessary and sufficient resources for the quantum-benefit RSP payoff}\label{iff}

To show that quantum coherence's static and dynamic resources are necessary and sufficient for demonstrating the quantum-benefit RSP payoff, $\mathcal{P}>0$, we first decompose the resulting received state of Bob's qubit $\rho$ into the diagonal and coherence parts by:
\begin{equation}
\rho=\rho_{d}+\rho_{c},\label{rho}
\end{equation}
for $\rho=\rho_{B|A},\rho_{B}$, where $\rho_{d}$ and $\rho_{c}$ are the matrices consisting of only the diagonal and off-diagonal (coherence) terms of $\rho$, respectively, concerning a given laboratory orthonormal basis $\{\ket{j}\}=\{\ket{0},\ket{1}\}$. Notably, $\rho_{c}$'s coherence characterizes the quantum mechanical feature of $\rho$. Now, to make such characteristic measurable, suppose that the state $\rho$ undergoes a quantum operation, $\mathcal{E}$. The coherence part becomes:
\begin{equation}
\mathcal{E}(\rho_{c})=\mathcal{E}(\rho)-\mathcal{E}(\rho_{d}). \label{opera}
\end{equation}
Further population measurements on the above states $\mathcal{E}(\rho)$ and $\mathcal{E}(\rho_{d})$ with respect to the basis $\{\ket{j}\}$ can be used to reveal the existence of the $\rho_{c}$'s coherence. Let us consider the following quantity: 
\begin{equation}
W_{\mathcal{E}}(\rho)= \bra{q}\mathcal{E}(\rho)\ket{q}-\sum_{j}\rho_{d}^{(jj)}\Omega_{qj}, \label{w}
\end{equation}
where $\ket{q}\in\{\ket{j}\}$, $\rho_{d}^{(jj)}=\bra{j}\rho_d\ket{j}$, and $\Omega_{qj}=\bra{q}\mathcal{E}(\ketbra{j})\ket{q}$. As shown and detailed in Ref.~\cite{li2012witnessing}, if $W_{\mathcal{E}}(\rho)$ is non-zero under a coherence creation operation $\mathcal{E}$, then $\rho$ possess a coherence part, $\rho_c$. That is, $\rho$ was created from a dynamic process involving the creation of quantum coherence. It is worth emphasizing that, in the RSP process, such a dynamic process of coherence creation is determined by Alice's measurement related to, $U$, and the state of the shared pair, $\rho_{AB}$, under Bob's accurate correction, $\hat{\pi}$. From the coherence's dynamic operation viewpoint, $U$ is a coherence creation operation and $\hat{\pi}$ is an operation of preserving quantum coherence. Therefore, the coherence detection~(\ref{w}) (or coherence witness) examines both the static and dynamic quantum coherence resources existing in the RSP process. See Appendix~\ref{c} for the implementation details of the coherence detection~(\ref{w}).

When considering the coherence examination in detail, $W_{\mathcal{E}}(\rho)>0$ implies that the strength of the $\rho$'s $\rho_c$-involved state transition to the state $\ket{q}$: $\bra{q}\mathcal{E}(\rho)\ket{q}$, is stronger than the strength of the $\rho_d$'s incoherent state transition to the state $\ket{q}$: $\sum_{j}\rho_{d}^{(jj)}\Omega_{qj}$. On the other hand, $W_{\mathcal{E}}(\rho)<0$ exhibits the population of $\ket{q}$ after the $\rho_c$-involved state transition can be smaller than that for $\rho_d$ without the coherence assisted in the state transition. Both cases reveal the coherence's characteristic of $\rho_c$: $W_{\mathcal{E}}(\rho)>0$ for the coherence-induced population increase, and $W_{\mathcal{E}}(\rho)<0$ for the coherence-induced population decrease, which are unattainable by the incoherent state $\rho_d$, using the coherence creation operation $\mathcal{E}$.

With $W_{\mathcal{E}}(\rho)$'s quantitive meaning of coherence, we use $W_{\mathcal{E}}(\rho)$ further to quantitatively examine the roles of Alice, Bob, and the state $\rho_{AB}$ in RSP. To show the coherence-induced enhancement via all the participants' inputs and $\rho_{AB}$ in the RSP, we consider the difference between $W_{\mathcal{E}_{U^{\dag}}}(\rho_{B|A})$ and $W_{\mathcal{E}_{U^{\dag}}}(\rho_{B})$ by Eq.~(\ref{w}) where $\ket{q}=\ket{0}$ and $\mathcal{E}=\mathcal{E}_{U^{\dag}}(\rho)= U^{\dag}\rho\ U$, and
\begin{equation}
U=\frac{1}{\sqrt{2}}\left[
\begin{matrix}
1 & 1 \\
e^{i\phi} & -e^{i\phi}
\end{matrix}\right]. \label{u}
\end{equation}
After a calculation, we arrive at the \textit{coherence enhancement} function:
\begin{eqnarray}
 \Delta W_{c,\mathcal{E}_{U^{\dag}}} &=& W_{\mathcal{E}_{U^{\dag}}}(\rho_{B\mid A})-W_{\mathcal{E}_{U^{\dag}}}(\rho_{B}) \nonumber\\
&=&-\frac{1}{2}[\rho_{32}+\rho_{23}+e^{2i\phi}\rho_{14}+e^{-2i\phi}\rho_{41}\label{dwcb} \\
&&\ \ \ \ \ -e^{i\phi}(\rho_{21}+\rho_{43})-e^{-i\phi}(\rho_{12}+\rho_{34})],\nonumber
\end{eqnarray}
for $ W_{\mathcal{E}_{U^{\dag}}}(\rho_{B\mid A})>0$. If $\Delta W_{c,\mathcal{E}_{U^{\dag}}}>0$, an experimental RSP process helps enhance the coherence-assisted transition to $\ket{q}$ [see also Eq.~(\ref{w})]. Since the coherence-induced population increase has been enhanced by RSP, we call $\Delta W_{c,\mathcal{E}_{U^{\dag}}}>0$ \textit{coherence enhancement} via RSP.

Notably, comparing Eq.~(\ref{dwcb}) with Eq.~(\ref{qbpayoff}), we find that:$\Delta W_{c,\mathcal{E}_{U^{\dag}}}$
\begin{equation}
\mathcal{P}=\Delta W_{c,\mathcal{E}_{U^{\dag}}},\label{dwc}
\end{equation}
where
\begin{eqnarray}
&&F(\rho_{B|A})\!=\!W_{\mathcal{E}_{U^{\dag}}}\!(\rho_{B\mid A})\!+\!\!\frac{1}{2},\!F(\rho_{B})\!=\!W_{\mathcal{E}_{U^{\dag}}}\!(\rho_{B})\!+\!\!\frac{1}{2}.\label{wb} 
\end{eqnarray}
Equations~(\ref{dwc}) and (\ref{wb}) imply that evaluation of the quantum-benefit RSP payoff~(\ref{qbpayoff}) can be exactly determined by measuring the coherence enhancement function~(\ref{dwcb}), where the state fidelity functions $F(\rho_{B|A})$ and $F(\rho_{B})$ are equivalent to the coherence detection, as shown in Eq.~(\ref{wb}). Therefore, an experiment shows the quantum benefit of RSP by $\mathcal{P}>0$ if and only if the experiment can demonstrate coherence enhancement, $\Delta W_{c,\mathcal{E}_{U^{\dag}}}>0$. Alternatively, as witnessed by the criterion: $\Delta W_{c,\mathcal{E}_{U^{\dag}}}>0$, the static and dynamic resources of quantum coherence are necessary and sufficient to gain the quantum-benefit RSP payoff, $\mathcal{P}>0$.

The above conclusion for a single run RSP is applicable to relationship between the mean quantum-benefit RSP payoff and the mean coherence enhancement defined below:
\begin{equation}
\begin{split}
\Delta W_{c,\text{avg}}=&\frac{1}{2\pi}\int_{0}^{2\pi} \Delta W_{c,\mathcal{E}_{U^{\dag}}}\ d\phi \\
=&-\frac{1}{2}(\rho_{23}+\rho_{32}), \label{eq:edwcb}
\end{split}
\end{equation}
which implies that
\begin{equation}
\mathcal{P}_{\text{avg}}=\Delta W_{c,\text{avg}},
\end{equation}
according to Eq.~(\ref{pay}). Therefore, $\mathcal{P}_{\text{avg}}>0$ if and only if $\Delta W_{c,\text{avg}}>0$.

\section{Coherence quantum benefits}\label{coherenceqb}

In Sec.~\ref{iff}, we show the required quantum resources for supporting quantum-benefit RSP payoff by using the coherence detection~(\ref{w}) underlying the introduced approach. While detecting coherence is equivalent to gaining the quantum-benefit RSP payoff (\ref{dwc}), the quantum operation $\mathcal{E}$ utilized in detecting coherence $\mathcal{E}_{\mathcal{E}_{U^{\dag}}}$~(\ref{u}) is a target-state-dependent coherence creation operation. $U$'s implementation depends on an RSP verifier's knowledge of $\ket{s}$ and the capability of creating a superposition state on the equator of the Bloch sphere. Therefore, if the receiver Bob plays the role of verifier to examine quantum benefits that can be gained in the experiment, such required capability of realizing $U$ goes against the original assumptions of limited Bob's ability. In the following, we will show that coherence-induced quantum benefits the verifier Bob can examine using a minimum of experimental resources.

As already pointed out, $W_{\mathcal{E}}(\rho)>0$ is the incoherent state that cannot mimic a kind of representation showing coherence under a chosen coherence creation operation $\mathcal{E}$. Therefore, given a verifier implementable coherence creation operation $\mathcal{E}$, we define the following criterion as a signature characteristic of quantum coherence powered by all the static and dynamic resources in an RSP, which we call the \textit{coherence quantum benefit} from the RSP process:
\begin{equation}
 \Delta W^>_{c,\mathcal{E}}\equiv W_{\mathcal{E}}(\tilde{\rho}_{B\mid A})-W_{\mathcal{E}}(\rho_{B}) >0,\label{cqb}
\end{equation}
for $W_{\mathcal{E}}(\tilde{\rho}_{B\mid A})>0$, where $\tilde{\rho}_{B\mid A}$ is the resulting experimental RSP state received by Bob without the assumption of perfect operations performed by Alice and Bob. Similarly, compared to the case of $W_{\mathcal{E}}(\rho)>0$, using the RSP process to enhance the coherence-induced population decrease, $W_{\mathcal{E}}(\rho)<0$, can also be utilized to characterize an RSP task. The following criterion also represents the coherence quantum benefit:
\begin{equation}
 \Delta W^<_{c,\mathcal{E}}\equiv W_{\mathcal{E}}(\rho_{B})-W_{\mathcal{E}}(\tilde{\rho}_{B\mid A}) >0,\label{w<}
\end{equation}
for $W_{\mathcal{E}}(\tilde{\rho}_{B\mid A})<0$. Thus, as experimental results satisfy either the criterion Eq.~(\ref{cqb}) or Eq.~(\ref{w<}), all the implementations in the experiment are identified as qualified to help Bob gain the coherence quantum benefit.

Since measuring either $\Delta W^>_{c,\mathcal{E}}$ or $\Delta W^<_{c,\mathcal{E}}$ requires implementing the coherence detection of $W_{\mathcal{E}}(\tilde{\rho}_{B\mid A})$ and $W_{\mathcal{E}}(\rho_{B})$, the experimental requirements additional to the original RSP required items are the population measurements in the basis $\{\ket{j}\}$ and the verifier's operation $\mathcal{E}$. Such needs are experimentally \textit{minimal} since only one measurement setting cannot reveal the coherence information from experimental results. This also indicates that when Bob serves as an RSP verifier, measuring state populations and performing a coherence creation operation become necessary.

\begin{figure*}[t] 
\centering
\includegraphics[width=1\textwidth]{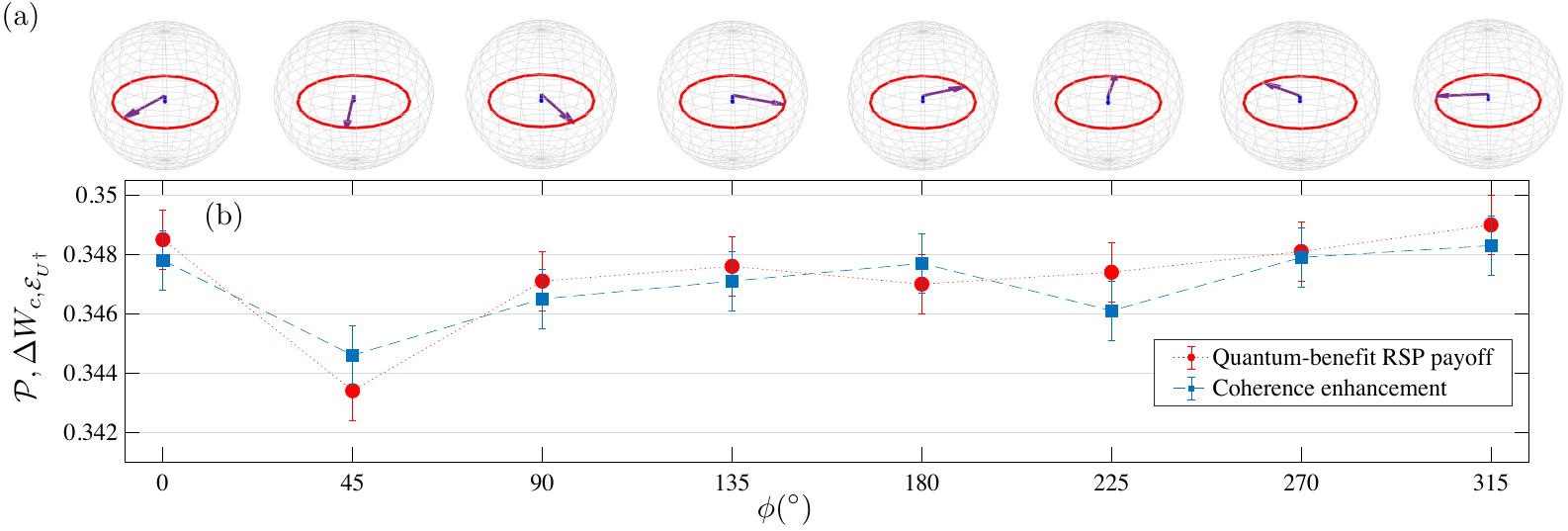}
\caption{Experimental quantum-benefit RSP payoff $\mathcal{P}$ and coherence enhancement $\Delta W_{c,\mathcal{E}_{U^{\dag}}}$. (a) Experimental Bob's states $\rho_{B|A}$ derived from the RSP (purple Bloch vector) and $\rho_{B}$ obtained without following the RSP protocol (blue Bloch vector) for eight different RSP tasks of $\ket{s}=(\ket{0}+e^{i\phi}\ket{1})/\sqrt{2}$. (b) As experimentally confirmed, the results of $\mathcal{P}$ and $\Delta W_{c,\mathcal{E}_{U^{\dag}}}$ for the eight RSP tasks of $\ket{s}$ are almost identical for all, which shows the required necessary and sufficient condition on coherence's static and dynamic resources for the RSP payoff. The experimental data used to calculate these values are summarized in Table~\ref{ifff}.}
\label{ifffg}
\end{figure*}

\begin{table}
\renewcommand\arraystretch{1.5}
\caption{Experimental data used to determine quantum-benefit RSP payoff $\mathcal{P}$ and coherence enhancement $\Delta W_{c,\mathcal{E}_{U^{\dag}}}$. Experimentally, $\mathcal{P}$ is determined by measuring the fidelity difference: $F(\rho_{B|A})-F(\rho_{B})$, where $F(\rho_{B|A})$ and $F(\rho_{B})$ are measured by using Bob's polarization analyzer with respect to the received states $\rho_{B|A}$ and $\rho_{B}$. Whereas $\Delta W_{c,\mathcal{E}_{U^{\dag}}}$ is obtained by performing two different kinds of coherence detection by: $W_{\mathcal{E}_{U^{\dag}}}(\rho_{B\mid A})-W_{\mathcal{E}_{U^{\dag}}}(\rho_{B})$. See Appendix~\ref{c} for the details of the coherence detection in our experiment.}
\begin{ruledtabular}
\begin{tabular}{ccccc}
 &$F(\rho_{B|A})$ &$F(\rho_{B})$ &$W_{\mathcal{E}_{U^{\dag}}}(\rho_{B\mid A})$ &$W_{\mathcal{E}_{U^{\dag}}}(\rho_{B})$\\ \hline
 $0^\circ$ &$0.8439$&$0.4954$  &$0.3489$ &$0.0011$ \\
$45^\circ$ &$0.8427$&$0.4993$  &$0.3463$ &$0.0017$ \\
$90^\circ$ &$0.8412$&$0.4941$  &$0.3457$ &$0.0012$ \\
$135^\circ$ &$0.8433$&$0.4957$  &$0.3486$ &$0.0015$ \\
$180^\circ$ &$0.8452$&$0.4982$  &$0.3488$ &$0.0011$ \\
$225^\circ$ &$0.8465$&$0.4991$  &$0.3464$ &$0.0013$ \\
$270^\circ$ &$0.8456$&$0.4975$  &$0.3491$ &$0.0012$ \\
$315^\circ$ &$0.8479$&$0.4989$  &$0.3494$ &$0.0011$ \\
\end{tabular}
\end{ruledtabular}\label{ifff}
\end{table}

In general, for the capability-limited Bob, his verifying operation $\mathcal{E}$ is not as generic as the operation $\mathcal{E}_{U^{\dag}}$ (\ref{u}). If $\mathcal{E}=\mathcal{E}_{U^{\dag}}$ in Eq.~(\ref{cqb}) and Alice and Bob perfectly implement the state measurement and correction, respectively, $\Delta W^>_{c,\mathcal{E}}$ becomes the coherence enhancement function (\ref{dwcb}). Compared to this extreme case, the resolution of showing coherence quantum benefits via either Eq.~(\ref{cqb}) or Eq.~(\ref{w<}) depends on all the experimental inputs and components that result in Bob's received state $\tilde{\rho}_{B\mid A}$, including the state of shared pair $\rho_{AB}$, all the operations by Alice and Bob, and the verifier Bob's operation $\mathcal{E}$. Therefore, this resolution reflects the performance of an experimental RSP task. In the following, we will first show how we experimentally perfomed the introduced RSP examinations with the quantum-benefit RSP payoff function~(\ref{qbpayoff}) and the coherence enhancement function~(\ref{dwcb}) in a single run RSP. Then we will present the experimental detection of the coherence quantum benefit according to the criteria~(\ref{cqb}) and~(\ref{w<}). 

\begin{figure*}[t] 
\centering
\includegraphics[width=1\textwidth]{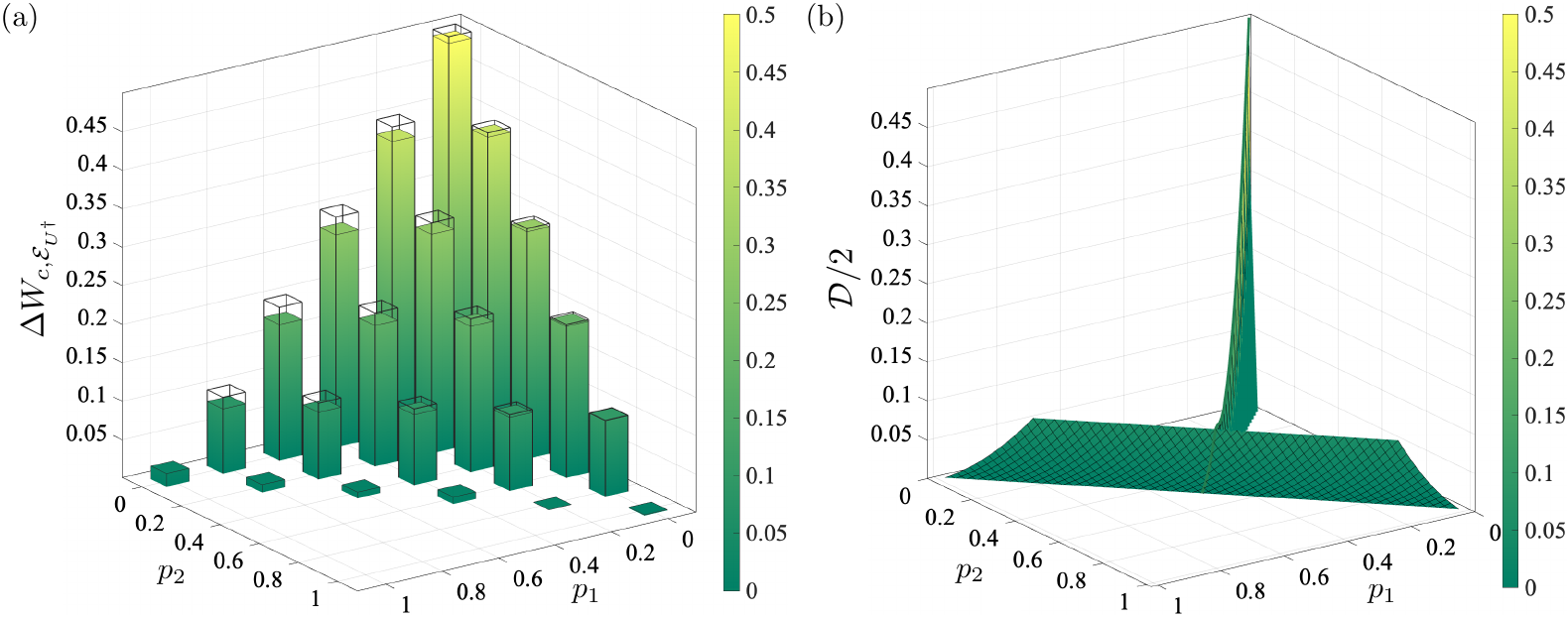}
\caption{Experimental coherence enhancement with different noisy shared states. (a) The experimental values of coherence enhancement vary with the noise intensities $p_1$ and $p_2$ in the shared pair states $\rho_{AB}=\rho_p$~(\ref{rhop}). Here we assume that the target state preparing for Bob is the state $\ket{s}$ for $\phi=0$, i.e., $\ket{+}$. The hollow bars with black frames represent the theoretical values of coherence enhancement. The experimental results are very close to the theoretical predictions. The experimental errors are small values of $\sim10^{-4}$ and therefore not shown therein. Compared to (b) based on the geometry quantum discord~\cite{dakic2012quantum} $\mathcal{D}$ to evaluation RSP, the method introduced in Ref.~\cite{dakic2012quantum} can only be applied to the noisy states with $p_{1}=p_{2}$ or Alice measures in the direction that parallels to the eigenvector corresponding to the largest eigenvalue of the shared state's correlation matrix.}
\label{3d}
\end{figure*}

\section{Experimental quantum-benefit RSP payoff and coherence enhancement}\label{expt}

In our experimental demonstration of verifying RSP process using the quantum-benefit RSP payoff function~(\ref{qbpayoff}) and the coherence enhancement function~(\ref{dwcb}), we created a bidirectionally pumped down-conversion source of polarization Sagnac interferometer (PSI) using a periodically polled KTP (ppKTP) crystal to generate polarization-entangled photon pairs~\cite{kim2006phase,fedrizzi2007wavelength,motazedifard2021nonlocal}. See~Fig.~\ref{exptset}(b). We observed $\sim10^4$ photon pairs per second and measured a state fidelity of $F\equiv\text{tr}(\rho_{\text{expt}}\!\ket{\psi^{-}}\!\!\bra{\psi^{-}})=(0.9917 \pm 0.0010)\%$ between the created state $\rho_{\text{expt}}$ and $\ket{\psi^{-}}$ for RSP. We then implement the RSP protocol with a high-quality EPR-pair source and precise realization of $U^\dag$. A detailed description of the technical parts is discussed in Appendix~\ref{a} for implementing the RSP protocol and Appendix~\ref{b} for the generation of the PSI entangled photon pairs.

\subsection{Necessary and sufficient coherence resources for RSP}\label{expt1}

To demonstrate that coherence resources are necessary and sufficient for gaining the quantum-benefit RSP payoff~(\ref{dwc}), we first experimentally implemented an RSP task as close as possible to an ideal standard RSP task. Our RSP verification shows that: $\mathcal{P}=0.4946 \pm0.0008$ and $\Delta W_{c,\mathcal{E}_{U^{\dag}}}=0.4914 \pm0.0009$, for the target state of $\ket{s}=\ket{+}=(\ket{0}+\ket{1})/\sqrt{2}$. These results are very close to the ideal value: $\mathcal{P}=\Delta W_{c,\mathcal{E}_{U^{\dag}}}=1/2$. Therefore, with the fact of $\mathcal{P}\approx\Delta W_{c,\mathcal{E}_{U^{\dag}}}$, this verification supports that coherence resources are necessary and sufficient to gain RSP benefits. The main steps for implementing the coherence detection are detailed in the Appendix~\ref{c}.

To illustrate RSP verification for different prepared states for Bob, we demonstrated that the quantum-benefit RSP payoff $\mathcal{P}$ is very close to the coherence enhancement $\Delta W_{c,\mathcal{E}_{U^{\dag}}}$ for different target states on the equator in Bob's Bloch sphere. Herein, we also consider the imperfection of $\rho_{AB}$ into account. We first created photon pairs with a state close to the state: $\rho_{AB}=\rho_{\text{noise}}=0.7\ket{\psi^{-}}\!\!\bra{\psi^{-}}+0.1\ket{00}\!\!\bra{00}+0.2\ket{11}\!\!\bra{11}$. The state fidelity of the created shared pairs and the state $\rho_{\text{noise}}$ is $(98.5\pm0.11)\%$. See Appendix~\ref{b} for how $\rho_{\text{noise}}$ was created in our experiment. As will be discussed in more detail in the following subsection, the reason of choosing such a state is to emphasize that our assessment approach can be applied to the shared photon pairs without any assumption compared to the existing verification methods.

Then, with the created $\rho_{AB}$ and the Alice and Bob's best inputs close to the ideal operations, our experimental results are shown in Fig.~\ref{ifffg}. First, as illustrated in Fig.~\ref{ifffg}(a), the $\rho_{B}$ without the RSP's help is very close to the sphere center, i.e., a maximally mixed state, while the red circle represents the contour of the states $\rho_{B|A}$ with purple Bloch vectors derived by following the RSP protocol. This increment from the center to the red circle represents the quantum beneficial effect for the RSP payoff $\mathcal{P}$~(\ref{qbpayoff}), which is equal to the coherence enhancement $\Delta W_{c,\mathcal{E}_{U^{\dag}}}$~(\ref{dwcb}), revealing the necessary and sufficient condition for RSP.

As shown in Fig.~\ref{ifffg}(b), $\mathcal{P}$ and $\Delta W_{c,\mathcal{E}_{U^{\dag}}}$ these two values are almost the same and positive for $8$ different target states $\ket{s}$ up to a very small errors of $\sim10^{-4}$. Theoretically, both the coherence detection and quantum-benefit payoff equals $0.35$ under the given noisy shared state. From the positive values of $\Delta W_{c,\mathcal{E}_{U^{\dag}}}$ and $\mathcal{P}$, one can confirm that all the inputs of Alice and Bob and the shared state $\rho_{\text{noise}}$ are qualified to gain the quantum benefits. Table~\ref{ifff} summarizes the measured data used to calculate the $\mathcal{P}$ and $\Delta W_{c,\mathcal{E}_{U^{\dag}}}$.

\begin{figure*}[t]
\centering
\includegraphics[width=0.96\textwidth]{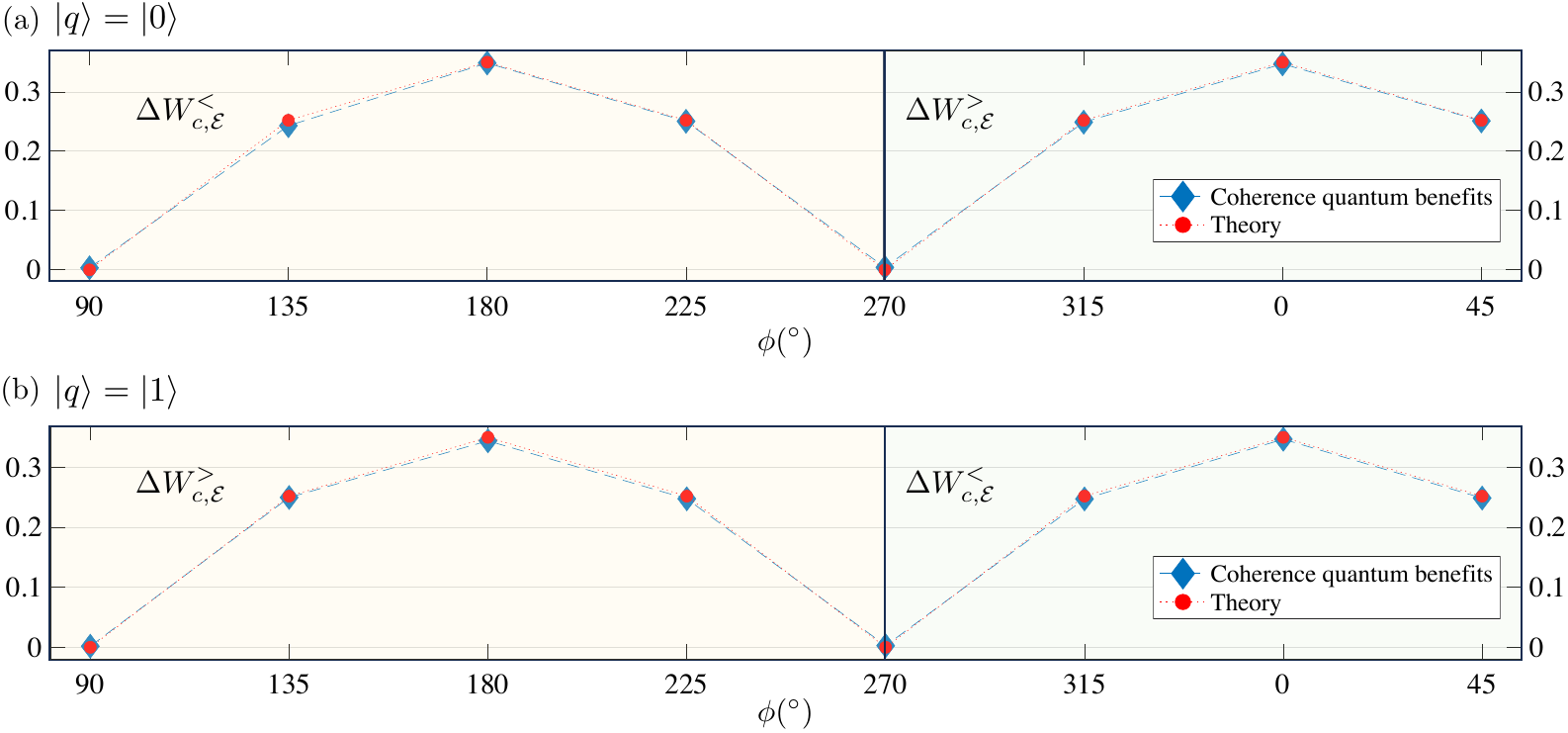}
\caption{Experimental coherence quantum benefits, $\Delta W^>_{c,\mathcal{E}}>0$ and $\Delta W^<_{c,\mathcal{E}}>0$. The coherence quantum benefits, $\Delta W^>_{c,\mathcal{E}}>0$ (\ref{cqb}) and $ \Delta W^<_{c,\mathcal{E}}>0$ (\ref{w<}), for (a) $\ket{q}=\ket{0}$ and (b) $\ket{q}=\ket{1}$, are experimentally determined according to the related coherence detection: $W_{\mathcal{E}}(\tilde{\rho}_{B\mid A})$ and $W_{\mathcal{E}}(\rho_{B})$. See Appendix~\ref{c} for the details of the coherence detection implemented in our experiment. As shown, since the six cases of verifying RSP for the six different states $\ket{s}=(\ket{0}+e^{i\phi}\ket{1})/\sqrt{2}$ reveal the coherence quantum benefits, for which either $\Delta W^>_{c,\mathcal{E}}>0$, or $\Delta W^<_{c,\mathcal{E}}>0$, these experimental RSP tasks are verified as qualified to show quantum benefits from the point of view of enhancing the quantum coherence. Whereas, for the RSP cases of $\phi=90^\circ,270^\circ$, we observed $\Delta W^>_{c,\mathcal{E}}=\Delta W^<_{c,\mathcal{E}}=0$, and can not infer existence of their quantum benefits.}
\label{blo}
\end{figure*}

\subsection{Coherence enhancement via RSP using different noisy shared photon pairs}\label{expt2}

To explicitly show that our verification can be applied to shared pairs without any state restriction, compared to the method introduced in Ref. \cite{dakic2012quantum}, which can only evaluate the RSP with the restricted shared isotropic states, we demonstrate by using anisotropic states for $\rho_{AB}$. Therefore, we assume the target shared state $\rho_{AB}$ incoherently consists of $\ket{\psi^{-}}$ and different intensities of noise:
\begin{equation}
\rho_{p}=(1-p_{1}-p_{2})\ket{\psi^{-}}\!\!\bra{\psi^{-}}+p_{1}\ket{00}\!\!\bra{00}+p_{2}\ket{11}\!\!\bra{11}, \label{rhop}
\end{equation}
where $0\leq p_{1}, p_{2}\leq1$ and $p_{1}+p_{2}\leq1$. See Appendix~\ref{b} for the details of experimentally creating the states $\rho_{p}$ under different noise intensities.

In the first experiment, we assume Alice aims to prepare the state $\ket{+}=(\ket{0}+\ket{1})/\sqrt{2}$ for Bob (i.e., $\phi=0$ for $\ket{s}$), with the shared pair of state $\rho_{AB}=\rho_{p}$. The faithful assessment results via coherence enhancement are shown in Fig.~\ref{3d}(a). In  Fig.~\ref{3d}(b), we show the states the method in Ref. \cite{dakic2012quantum} can verify. Their approach can only evaluate performance for the states with $p_{1}=p_{2}$, or the eigenvector of the correlation matrix turns to parallel to Alice's measurement direction. The experimental result shows that our approach faithfully evaluates the performance of RSP in terms of coherence enhancement, revealing the properties of the shared state.

\section{Experimental coherence quantum benefits}\label{expt}

To show how the coherence quantum benefits, $\Delta W^>_{c,\mathcal{E}}>0$ (\ref{cqb}) and $ \Delta W^<_{c,\mathcal{E}}>0$ (\ref{w<}), are helpful to efficiently verify experimental RSP tasks, we first assume that Bob only have limited ability to perform $\mathcal{E}$ on the experimental state $\tilde{\rho}_{B|A}$, and suppose that he can only realize a Hadamard transformation $\mathcal{E}_{H}(\rho)=H^\dag\rho H$, where $H=\ketbra{+}{0}+\ketbra{-}{1}$. The created photon pairs close to the state, $\rho_{AB}=\rho_{\text{noise}}$, are utilized in the experimental RSP. We consider 8 different target states prepared for Bob. As shown in Fig.~\ref{blo}, the coherence quantum benefits vary with the states close to the target states on the equator. There are two cases: $\phi=90^\circ,270^\circ$, for which the RSP performance can not be seen via the coherence quantum benefits for $\Delta W^>_{c,\mathcal{E}}=\Delta W^<_{c,\mathcal{E}}=0$.

Other cases are identified as qualified to show coherence quantum benefits. The theoretical and experimental average coherence quantum benefits over the whole equator is $0.22$ and $0.21$, respectively. All the experimental results are very close to the theoretical predictions within an uncertainty of $\sim10^{-4}$, showing that the RSP tasks can be efficiently evaluated under the minimum experimental requirements as predicted.

\section{Conclusion and outlook}\label{co}

Preparing remote states has played a vital role in quantum information processing for resource-efficient advantages over quantum teleportation. While appraising the remote state preparation (RSP) performance is essential when considering practical implementations, the existing evaluating methods require the optimization of the RSP protocol, where the sender and receiver need the ability to perform quantum state tomography assisted by two-way communication. This goes against the original simple setting for RSP. 

We propose a new approach to assess the RSP task based on the quantum benefits powered by quantum coherence's static resources of the shared pairs and the dynamic resources both the RSP participants input. From the task-oriented viewpoint to consider the quantum benefits, our approach shows that these coherence resources play a necessary and sufficient role in the payoff for RSP. Notably, from the point of view of enhancing the quantum coherence via RSP to consider the quantum benefits, it provides the most efficient RSP assessment, requiring only the receiver's minimum of one additional coherence operation to verify RSP. 

Furthermore, we experimentally demonstrate the introduced RSP assessment with different photon pair states generated from a high-quality polarization Sagnac interferometer. Our results support the existence of the necessary and sufficient role played by the static and dynamic quantum coherence resources. They also show that efficient RSP verification is experimentally feasible using the minimum experimental settings of manipulating photon polarization states. The introduced concept and methods may be applicable to investigate efficient verification of measurement-based quantum information processes and their underlying static and dynamic resources in practical quantum information scenarios with foundational principles close to RSP, such as one-way quantum computing \cite{raussendorf2001one,you2007efficient,tanamoto2009efficient,wang2010robust} and entanglement-enabled networks \cite{pirker2018modular,wehner2018quantum}.

\section*{Acknowledgements}

This work was partially supported by the National Science and Technology Council, Taiwan, under Grant Numbers NSCT 111-2123-M-006-001, NSCT 112-2123-M-006001.

\appendix

\section{Implementation of RSP protocol}\label{a}

We realize the RSP protocol with respect to a target state, $U\ket{s_0}=(\ket{0}+e^{i\phi}\ket{1})/\sqrt{2}$. The main steps for implementing the RSP protocol are as follows.

First, Alice performs the operation $U^{\dag}$ such that $\ket{\psi^{-}}$ becomes
\begin{equation}
U^\dag\ket{\psi^{-}}=\frac{1}{\sqrt{2}}(\ket{s_{0}}\otimes U\ket{s_{0}^{\bot}}-\ket{s_{0}^{\bot}}\otimes U\ket{s_{0}}).\label{explain1}
\end{equation}
In our experimental RSP (Fig.~\ref{exptset}), $U^\dag$ was realized by using a setting composed of QWP, HWP, and QWP (QHQ) and a HWP in the polarization analyzer (See Appendix~\ref{c}).

Second, Alice's qubit is measured in the basis $\{\ket{s_{0}},\ket{s_{0}^{\bot}}\}=\{\ket{0},\ket{1}\}$. Experimentally, it was done by the polarization analyzer in mode $A$ in the experimental setup. The PBS in the polarization analyzer helps distinguish the states $\ket{0}$ and $\ket{1}$. Her measurement results imply the following Bob's qubit states in the ideal case of Eq.~(\ref{explain1}): $\ket{0}\rightarrow U\ket{1}$ and $\ket{1}\rightarrow U\ket{0}$. For practical cases in experimental implementations, the $U$ in the above relationship is replaced by a imperfect quantum operation: $\ket{0}\rightarrow\tilde{\mathcal{E}}_U(\ketbra{1})$ and $\ket{1}\rightarrow\tilde{\mathcal{E}}_U(\ketbra{0})$.

Third, according to the RSP protocol, the resulting state received by Bob via the RSP protocol $\tilde{\rho}_{B\mid A}$ is composed of the part without Bob's correction and the part required Bob's $\hat{\pi}$ operation:
\begin{equation}
\tilde{\rho}_{B\mid A}=p_A(1)\tilde{\mathcal{E}}_{U} \left(\ket{0}\!\!\bra{0}\right)+p_A(0)\mathcal{E}_{\hat{\pi}}\circ\tilde{\mathcal{E}}_{U} \left(\ket{1}\!\!\bra{1}\right),
\end{equation}
where $p_A(n)$ for $n=0,1$ denotes the probabilities that Alice obtain the results $n$, $\mathcal{E}_{\hat{\pi}}$ denotes the experimental implementation of $\hat{\pi}$. Therefore, the experimentally created state $\tilde{\rho}_{B\mid A}$ by following the RSP protocol can be examined by using the proposed approach detailed in the main text.

\section{Implementation of coherence detection}\label{c}

As shown in Secs.~\ref{iff}~and~\ref{coherenceqb}, Bob's coherence creation operation $\mathcal{E}$ can be the minimum additional experimental setting required to very RSP process. To realize the $\mathcal{E}$, we first use QHQ plays the phase shifter, where the first QWP sets the angle at $+45^\circ$ with respect to the fast axis, and the second QWP set as $-45^\circ$. If the middle HWP's angle is set at $2\theta_{\phi_b}$, the QHQ phase shifter's operation is described by:
 \begin{equation}
 \left[\begin{matrix}
 1 & 0 \\
0 & e^{i\theta_{\phi_b}}
 \end{matrix}\right].
 \end{equation}
If combining of QHQ and the following HWP (i.e., the one used in a polarization analyzer), which is set at $22.5^\circ$, the verifying operation $\mathcal{E}=\mathcal{E}_{U^{\dag}}$~(\ref{u}) can be experimentally realized. When $2\theta_{\phi_b}=0$, the operation $\mathcal{E}$ is equal to Hadamard transformation. Bob's correction $\hat{\pi}$ is implemented by QHQ in the same manner. Similarly, Alice's $U^\dag$ can be realized in the same manner as follows. The two QWPs are set as $\pm 45^\circ$, either. The middle HWP's angle $2\theta_{\phi_a}$ of phase shifter set corresponding to the required target state $U\ket{s_0}$.

Implementing the coherence detection~(\ref{w}) consists of measuring the $\bra{q}\mathcal{E}(\rho)\ket{q}$ and $\sum_{j}\rho_{d}^{(jj)}\Omega_{qj}$ by the verifier Bob, where $\rho_{d}^{(jj)}=\bra{j}\rho_d\ket{j}$, and $\Omega_{qj}=\bra{q}\mathcal{E}(\ketbra{j})\ket{q}$. The main steps for the measurements of these two quantities are summarized as follows.

First, to measure $\bra{q}\mathcal{E}(\rho)\ket{q}$, at the beginning, Alice uses QHQ at the angle as $(+45^\circ, 2\theta_{\phi a}, -45^\circ)$ to implement $U^\dag$ to prepare the target state $\rho$ for Bob. Bob uses QHQ at the angle as $(+45^\circ, 2\theta_{\phi b}, -45^\circ)$ to implement $\mathcal{E}$ on the received state $\rho$. Then he uses the polarization analyzer to measure the resulting population $\ket{q}$ of $\mathcal{E}(\rho)$.

Second, to measure $\rho_{d}^{(jj)}=\bra{j}\rho_d\ket{j}$, Alice uses QHQ at the angle as $(+45^\circ, 2\theta_{\phi a}, -45^\circ)$ while Bob set both QWPs in QHQ at $0^\circ$ and the HWP as $2\theta_{\phi b}=0$. He then measures the $\rho_{d}^{(jj)}$ by the polarization analyzer, respectively.

Third, to measure $\Omega_{qj}=\bra{q}\mathcal{E}(\ketbra{j})\ket{q}$, Alice uses QHQ at the angle as $(0^\circ, 0^\circ, 0^\circ)$ to prepare the state $\ket{j}$ for Bob. Bob follows the same method as the first step to implement $\mathcal{E}$ and uses the polarization analyzer to measure the resulting population $\ket{q}$ of $\mathcal{E}(\ketbra{j})$.

Finally, with the results obtained above, Bob obtains a result of coherence detection~(\ref{w}).

\section{PSI entangled photon source}\label{b}

As illustrated in Fig.~\ref{exptset}(b), a CW laser at $405$ nm serves as the pumping source for our PSI entangled photon source. After reflecting by a long-pass dichroic mirror ($\text {DM}$), the laser beam then pumps the PSI which consists of a $\text {dPBS}$, a $\text {dHWP}$ set at $\pi/4$ and two mirrors ($\text {M}_{2}$ and $\text {M}_{3}$). The ppKTP crystal of 25 mm long is mounted at the center of PSI. We keep the ppKTP crystal at $27.1\pm 0.01^{\circ}$C with a temperature controller and a thermoelectric cooler. The laser beam, which is reflected at ${\text{dPBS}}$, i.e., the vertical polarization component, is rotated by ${\text {dHWP}}$ to horizontal polarization and then incident into the ppKTP crystal. The down-conversion process generates orthogonal polarization photon pairs, $\ket{H_{s}}_{\circlearrowleft}$
and $\ket{V_{i}}_{\circlearrowleft}$ of which the wavelength is 810 nm, where $s$ and $i$ are respectively denoted as signal and idler photon. Here, the subscripted rotation symbol denotes the spatial mode of the photons propagating in the counterclockwise direction in the PSI.

The horizontal polarization component of the pumping laser beam is transmitted at ${\text {dPBS}}$ and propagates in the clockwise direction in the PSI. Then, 810 nm photon pairs $\ket{H_{s}}_{\circlearrowright}$ and $\ket{V_{i}}_{\circlearrowright}$ are generated by the ppKTP crystal. After reflection on ${{\text{M}_{3}}}$, $\ket{H_{s}}_{\circlearrowright}$ and $\ket{V_{i}}_{\circlearrowright}$ are rotated to $\ket{V_{s}}_{\circlearrowright}$ and $\ket{H_{i}}_{\circlearrowright}$ by the ${\text {dHWP}}$. As the two oppositely propagating photon pairs go through the $\text {dPBS}$, where the vertical photons are reflected, and horizontal photons are transmitted out of $\text {dPBS}$, the $\ket{H_{s}}_{\circlearrowleft}$ and $\ket{V_{s}}_{\circlearrowright}$ pass to the path A as $\ket{H}_{A}$ and $\ket{V}_{A}$, and $\ket{V_{i}}_{\circlearrowleft}$ and $\ket{H_{i}}_{\circlearrowright}$ pass to the path B as $\ket{V}_{B}$ and $\ket{H}_{B}$. 

To generate polarization-entangled photon pairs under the PSI source, we set the angle of $\text{HWP@405 nm}$ at -$\pi/8$, $\text{QWP@405 nm}$ at 0, and a rotation angle of $\varphi=-15.4^{\circ}$ around the fast axis of $\text{QWP@405 nm}$. The resulting states generated from the PSI is close to the target state of $\ket{\psi^{-}}=(\ket{H}_A\ket{V}_B-\ket{V}_A\ket{H}_B)/\sqrt{2}$, where $\ket{H}_k$ and $\ket{V}_k$ denote the horizontal and vertical polarization states in the path $k$ for $k=A,B$, respectively. We use two DMs placed on each path to reflect the 405 nm laser beam to avoid collecting incorrect photons. In each path, a QHQ setting composed of one $\text{HWP}$ and two $\text{QWP}$ is used to ether implement Alice's $U\dag$ or Bob's coherence creation operation $\mathcal{E}$. The following ${\text {HWP}}$, ${\text {QWP}}$, and ${\text {PBS}}$ form a polarization analyzer used to measure the photon pairs. The 810 nm BF and LPF select photons with the corrected wavelength. The photon pairs are collected by couplers and sent to single photon counting modules via single-mode fibers. A FGPA-based coincidence processing unit is used to measure the coincidence count rates under a 4.4 ns coincidence window. We detected $10^4$ of entangled photon pairs per second at $5$ mW.

To realize the noisy states: $\rho_{\text{noise}}$ in Sec.~\ref{expt1} and $\rho_{p}$ (\ref{rhop}) in Sec.~\ref{expt2}, we experimentally create the states for RSP with respect to the state~(\ref{rhop}) by setting different duration times for generating the $\ket{\psi^-}\!\!\bra{\psi^-}$, $\ket{00}\!\!\bra{00}$ and $\ket{11}\!\!\bra{11}$. The generating method of $\ket{\psi}\bra{\psi}$ is as described in the previous section. To realize the states $\ket{00}\!\!\bra{00}$ and $\ket{11}\!\!\bra{11}$, we set the HWP@405 at $0^\circ (45^\circ)$ for preparing the state $\ket{H} (\ket{V})$ to pump into the PSI, generating the separate state $\ket{V}_A\ket{H}_B (\ket{H}_A\ket{V}_B)$. After going through the first HWP at $45^\circ$ in path A, we obtain $\ket{H}_A\ket{H}_B (\ket{V}_A\ket{V}_B)$ corresponding to $\ket{00}\!\!\bra{00}$ $(\ket{11}\!\!\bra{11})$. We take $\rho_{\text{noise}}$ where $p_1=0.1$, $p_2=0.2$ as a concrete example. Since the ratio of the states involving $\ket{\psi^-}\!\!\bra{\psi^-}$, $\ket{00}\!\!\bra{00}$ and $\ket{11}\!\!\bra{11}$ is:
\begin{equation}
\ket{\psi^-}\!\!\bra{\psi^-}:\ket{00}\!\!\bra{00}:\ket{11}\!\!\bra{11}=7:1:2,
\end{equation}
the ratio of the state creation duration times is $7:1:2$ as well.


\begin{thebibliography}{12}
\bibitem{corre}Correlated coherence is a measure of coherence with the local part removed denoted as $C_{cc}(\rho_{AB})=C(\rho_{AB})-C(\rho_{A})-C(\rho_{B})$, where $C$ is the measurement of coherence.
\end{thebibliography}
\end{document}